\documentclass[prb,amsmath,amssymb]{revtex4}
\usepackage{graphicx}% Include figure files
\usepackage{dcolumn}% Align table columns on decimal point
\usepackage{bm}% bold math

\newcommand{\mapright}[1]{\smash{\mathop{\hbox to 1.0cm{\rightarrowfill}}\limits^{#1}}}

\begin{document}

\preprint{}

\title{Quasiparticle Dissipation in d-wave Superconductor Phase Qubit}% Force line breaks with \\

\author{S. Kawabata$^{1}$\footnote{Electronic address: s-kawabata@aist.go.jp}, S. Kashiwaya$^2$, Y. Asano$^3$, and Y. Tanaka$^4$}
%\email{s-kawabata@aist.go.jp}
\affiliation{%
$^1$Nanotechnology Research Institute and Research Consortium for 
Synthetic Nano-Function Materials Project (SYNAF), National Institute of 
Advanced Industrial Science and Technology (AIST), 1-1-1 Umezono, Tsukuba, 
Ibaraki 305-8568, Japan \\
$^2$Nanoelectronics Research Institute, AIST, 1-1-1 Umezono, Tsukuba, 
Ibaraki 305-8568, Japan \\
$^3$Department of Applied Physics, Hokkaido University,
Sapporo, 060-8628, Japan\\
$^4$Department of Applied Physics, Nagoya University,
Nagoya, 464-8603, Japan
}%

\date{\today}

\begin{abstract}
Recent phase sensitive experiments on high $T_c$ superconductors, e.g., YBa$_2$Cu$_3$O$_7$ single crystals, have established the $d$-wave nature of the cuprate materials. 
Here we discuss how to make use of $d$-wave Josephson junctions in the construction of a phase qubit.
We especially focus on the effect of quasiparticle dissipation on the macroscopic quantum tunneling which corresponds to the measurement  process for the $d$-wave phase qubit.
\end{abstract}

%\pacs{74.50.+r, 03.65.Yz, 05.30.-d}% PACS, the Physics and Astronomy
\maketitle

Since macroscopic systems are inherently dissipative, there arises a fundamental question of what is the effect of dissipation on the macroscopic quantum tunneling (MQT).
This issue was solved by Caldeira and Leggett in 1981 by using the path-integral method and they showed that MQT is depressed by dissipation.~\cite{rf:CL81}
This effect has been verified in experiments on $s$-wave Josephson junctions shunted by an Ohmic normal resistance).~\cite{rf:Cleland88}
As was mentioned by Eckern {\it et. al.}, the influence of the quasiparticle dissipation is quantitatively weaker than that of the Ohmic dissipation on the shunt resistor.~\cite{rf:Eckern84}
This is due to the existence of an energy gap $\Delta$ for the quasiparticle excitation in superconductors.
Therefore, in an ideal $s$-wave Josephson junction without the shunt resistance, the suppression of the MQT rate due to the quasiparticle dissipation is very weak.
In the $d$-wave superconductors, on the other hand, the the gap vanishes in certain directions,~\cite{rf:d-wave1} hence quasiparticles can be excited even at sufficiently low temperature regime.
Therefore, the nodal quasiparticles give a crucial contribution to the MQT.

In this paper, we will discuss how to make use of $d$-wave Josephson junctions in the construction of a phase qubit (Fig. 1).~\cite{rf:Kawabata}
The measurement of the qubit state utilizes the escape from the cubic potential via MQT. To measure the occupation probability $P_1$ of state $| 1 \rangle$, we pulse microwaves at frequency $\omega_{12}$, driving a $1\to 2$ transition. The
large tunneling rate $\Gamma_2$ then causes state $| 2 \rangle$ to rapidly
tunnel. 
After tunneling, the junction behaves as an open
circuit, and a dc voltage of the order of the superconducting
gap appears across the junction. This voltage
is readily measured with a room-temperature
amplifier. Thus the occupation probability $P_1$ is equal to
the probability of observing a voltage across the junction
after the measurement pulse.

\begin{figure}[h]
\includegraphics[width=14.0cm]{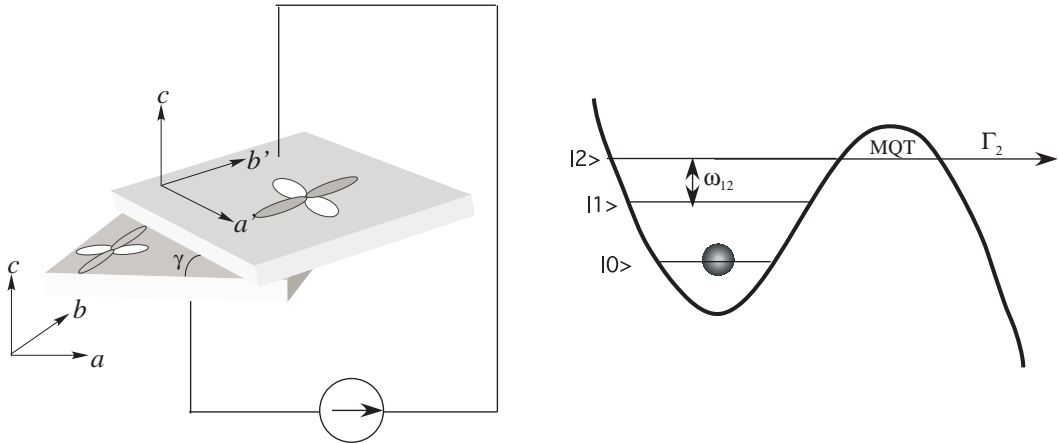}
\caption{Schematic drawing of the $d$-wave phase qubit. $\gamma$ is the twist angle about the $c$-axis.  }
\label{fig1}
\end{figure}
\section{Theory}

In this section, we will show the calculation of the effective action and the MQT rate for the $d$-wave phase qubit (the current biased $c$-axis twist Josephson junction~\cite{rf:Takano02,rf:Maki03}).
In Fig.1, we show schematic of this junction.
In this figure, $\gamma$ is the twist angle about the $c$-axis ($0 \le \gamma \le \pi/4$). 
Such a junction was recently fabricated by using the single crystal whisker of Bi${}_2$Sr${}_2$CaCu${}_2$O${}_{8+\delta}$.
Takano $et. al.$ measured the twist angle dependence of the $c$-axis Josephson critical current and showed a clear evidence of the $d_{x^2-y^2}$ symmetry of the pair potential.~\cite{rf:Takano02}
In the following, we assume that the tunneling between the two superconductors is described in terms of the coherent tunneling ($\left| t(\mbox{\boldmath $k$},\mbox{\boldmath $k$}')\right|^2 
 = 
 \left| t \right|^2 \delta_{\mbox{\boldmath $k$}_{\parallel},\mbox{\boldmath $k$'}_{\parallel}}
$).
For simplicity, we also assume that each superconductors consist of single  CuO${}_2$ layer, $\Delta_1 (\mbox{\boldmath $k$})=\Delta_0 \cos 2 \theta$, and $\Delta_2 (\mbox{\boldmath $k$})=\Delta_0 \cos 2 \left( \theta + \gamma \right)$.
Moreover, we consider the low temperature limit ($k_B T \ll \Delta_0$).

By using the functional integral method, the ground partition function for the system can be written as follows
\begin{eqnarray}
Z
= 
\int 
D \phi (\tau) 
\exp
\left[
  - \frac{S_{\mathrm{eff}}[\phi]}{\hbar} 
\right]
,
\end{eqnarray}
where $\phi=\phi_1-\phi_2$ is the phase difference across the junction.
In this equation, the effective action is given by 
\begin{eqnarray}
S_{\mathrm{eff}}[\phi]
&= &
\int_{0}^{\hbar \beta} d \tau 
\left[
   \frac{M}{2} 
   \left(
   \frac{\partial \phi ( \tau) }{\partial \tau}
   \right)^2
   + 
   U(\phi)
\right]
+
S^{[\alpha]}[\phi]
,
\nonumber\\
\\
S^{[\alpha]}[\phi]
&= &
-
\int_{0}^{\hbar \beta}  d \tau 
 \int_{0}^{\hbar \beta} d \tau'
  \alpha (\tau - \tau') \cos \frac{\phi(\tau) - \phi (\tau') }{2}
  ,
\end{eqnarray}
where $M=C(\hbar/2e)^2$ is the mass and $ U(\phi) $ is the tiled washboard potential
\begin{eqnarray}
 U(\phi) 
 = 
  -E_J(\gamma) \left(  \cos \phi +  \frac{I_{\mathrm{ext}}}{I_C(\gamma) } \phi \right)
  .
\end{eqnarray}
In this equation, $E_J=\left(  \hbar/2 e \right) I_C$ is the Josephson coupling energy, 
$I_C$ is the Josephson critical current, and $I_{\mathrm{ext}}$ is the external current.

In the following, we will consider the effect of the nodal quasiparticles on MQT.
For this purpose, we first calculate the dissipation kernel $\alpha (\tau)$ for two types of the $c$-axis junction, $i.e$., (1) $\gamma=0$ and (2) $\gamma \ne 0$ (here we will show the result for $\gamma = \pi/8$ case only.).

In the case of the $c$-axis junction with $\gamma=0$, the nodes of the pair potential in the two superconductors are in the same direction.
Therefore, the node-to-node quasiparticle tunneling is possible even at very low temperatures.
In this case, the asymptotic form of the dissipation kernel at the zero temperature is given by 
\begin{eqnarray}
\alpha(\tau) 
\approx
\frac{3 \hbar^2  | t|^2 N_0^2 }{16 \Delta_0}  \frac{1}{|\tau|^3}
\end{eqnarray}
for $\Delta_0 |\tau| /\hbar \gg1$.
This gives the super-Ohmic Dissipation.

On the other hand, in the case of the finite twist angle ($\gamma=\pi/8$), the asymptotic behavior of the dissipation kernel is given by an exponential function due to the suppression of the node-to-node quasiparticle tunneling, $i.e.,$
\begin{eqnarray}
\alpha(\tau)
\sim
 \exp \left( - \frac{1}{\sqrt{2}} \frac{\Delta_0 |\tau| }{\hbar} \right)
\end{eqnarray}
for $\Delta_0 |\tau| /\hbar \gg1$.

The MQT rate at the zero temperature is given by~\cite{rf:MQT1}

\begin{eqnarray}
\Gamma
=
\lim_{\beta \to \infty} \frac{2}{\beta} \mbox{ Im}\ln Z
\approx
A
\exp \left( - \frac{S_B}{ \hbar} \right),
\end{eqnarray}
where $S_B= S_{\mathrm{eff}}[\phi_B]$ is the bounce exponent, that is the value of the the action $S_{\mathrm{eff}}$ evaluated along the bounce trajectory $\phi_B(\tau)$.
Using the instanton method, we obtain the analytic expressions for the MQT rate:
\begin{eqnarray}
\thinspace
\frac{\Gamma(0)}{\Gamma_0(0)}
&\approx&
 \exp \left[ 
 -B(0)
 - 0.14
  \frac{ \hbar I_C(0)}{\Delta_0^2}
  \sqrt{ \frac{\hbar}{2 e} \frac{I_C(0)}{C}}
 \left\{
  1 -
  \left(
  \frac{I_{\mathrm{ext}}}{I_C(0)}
  \right)^{2}
\right\}^{5/4}
  \right]
,
\\
\frac{\Gamma(\pi/8)}{\Gamma_0(\pi/8)}
&\approx&
 \exp 
 \left[
    - B\left(\pi/8 \right)
\right]
,
\nonumber\\
\end{eqnarray}
where,
\begin{eqnarray}
B(\gamma)
=
\frac{12}{5e} \sqrt{ \frac{\hbar}{2 e} I_C(\gamma) C}
  \left(
                \sqrt{1+ \frac{\delta M(\gamma)}{M} } -1
    \right)
	\left\{
1-
    \left(
              \frac{I_{\mathrm{ext}}}{I_C(\gamma)}
    \right)^{2}
	\right\}^{5/4}    ,
\end{eqnarray}
and $\Gamma_0(\gamma)$ is the decay rate without the quasiparticle dissipation.
In eq.(10), $\delta M(\gamma)$ is the renormalization mass.
As an example, for $\Delta_0=42.0$ meV,~\cite{rf:Pan01} $I_c(\gamma=0)=1.45 \times10^{-4}$ A, $C=10$ fF, and $I_{\mathrm{ext}}/I_C(\gamma) =0.9$, we obtain
\begin{eqnarray}
\frac{\Gamma(\gamma)}{\Gamma_0(\gamma)}
\approx
\left\{
\begin{array}{rl}
90 \ \% & \quad \mbox{for} \quad \gamma=0 \\
96 \ \% & \quad\mbox{for} \quad \gamma=\pi/8
\end{array}
\right.
.
\end{eqnarray}
Therefore, the node-to-node quasiparticle tunneling in the case of the $\gamma=0$ junction gives rise to large reduction of the MQT rate in compared with the $\gamma=\pi/8$ case.

\section{Summary}
To summarize, we have investigated the effect of quasiparticle dissipation for the d-wave phase qubit ($c$-axis twist Josephson junction).
Within the coherent tunneling approximation, we find  the super-Ohmic dissipation in the case of the zero twist angle qubit.
This dissipation is caused by the node-to-node quasiparticle tunneling between the two superconductors.
Therefore, in this case, the fidelity of the readout process is very low.
On the other hand, in the case of the finite twist angle qubit, the suppression of the MQT rate is very weak in compared with the $\gamma = 0$ case due to the inhibition of the node-to-node quasiparticle tunneling.
This gives the high-fidelity readout for the $d$-wave phase qubit.

% \appendix

% \section{}
% \label{}

\end{document}